\documentclass[twocolumn,twoside,letterpaper,11pt]{article}
\usepackage{graphicx,xrrc,natbib}
\usepackage{times}

\def\ltsim{\hbox{\raise 2pt \hbox {$<$} \kern-1.1em \lower 4pt \hbox {$\sim$}}}
\def\ltapprox{\hbox{\raise 2pt \hbox {$<$} \kern-1.1em \lower 5pt \hbox 
{$\approx$}}}
\def\gtsim{\hbox{\raise 2pt \hbox {$>$} \kern-1.1em \lower 4pt \hbox {$\sim$}}}
\def\gtapprox{\hbox{\raise 2pt \hbox {$>$} \kern-1.1em \lower 5pt \hbox 
{$\approx$}}}

\def\arcsec{$^{\prime\prime}$}
\def\arcmin{$^{\prime}$}

\begin{document}

\title{RADIO OBSERVATIONS OF CLUSTER MERGERS}
\author{ Luigina Feretti                        } 
\institute{ Istituto di Radioastronomia CNR/INAF      } 
\address{   Via P. Gobetti 101, 40129 Bologna, ITALY        } 
\email{     lferetti@ira.cnr.it                         } 

\maketitle

\abstract{ 
A very important aspect of the radio emission from galaxy
clusters is represented by the diffuse radio sources associated
with the intracluster medium: radio halos and relics.  These radio
sources indicate the existence of large scale magnetic fields and of a
population of relativistic electrons in the cluster volume.  The
observational results provide evidence that these phenomena are
related to cluster merger activity, which supplies the energy to the
reacceleration of the radiating particles.  The details of the
halo-merger connection are investigated through the comparison between
the radio and the X-ray emission.  }
 
\section{Introduction}

Clusters of galaxies are the largest gravitationally bound systems in
the Universe.  Most of the gravitating matter in any cluster is in the
form of dark matter ($\sim$ 80\%). Some of the luminous matter is in
galaxies ($\sim$ 3-5\%), the rest is in diffuse hot gas ($\sim$
15-17\%), detected in X-ray through its thermal bremsstrahlung
emission.  This thermal plasma, consisting of particles of energies of
several keV, is commonly referred to as Intracluster Medium (ICM).  In
recent years it has become clear that the ICM can also contain highly
relativistic particles, with energy density $<$1\% than the energy
density in the thermal plasma, but which are very important in the
cluster formation and evolution.

We do not know yet how common this non-thermal 
component is in all clusters.  The
most detailed studies of this component come from the radio
observations.  A number of clusters of galaxies is known to contain
large-scale diffuse radio sources which have no obvious connection
with the cluster galaxies, but are rather associated with the ICM.
These sources are classified in two groups, {\it radio halos} and 
{\it relics},
according to their location at the cluster center or cluster
periphery, respectively.  The synchrotron origin of the emission from
these sources requires the presence of cluster-wide magnetic fields of
the order of $\sim$ 0.1-1 $\mu$G, and of a population of relativistic
electrons with Lorentz factor $\gamma >>$ 1000 and energy density of
$\sim$ 10$^{-14}$-10$^{-13}$ erg cm$^{-3}$.

Clusters are formed by hierarchical structure formation processes.  In
this scenario, smaller units formed first and merged to larger and
larger units in the course of time.  The merger activity appears to be
continuing at the present time, and explains the relative abundance of
substructure and temperature gradients detected in Abell clusters by
the X-ray observations.  The ICM in merging clusters is likely to be
in a violent or turbulent dynamical state.  It is found that the
diffuse radio sources are detected in clusters which have recently undergone
a merger event, thus leading to the idea that they are energized by
turbulence and shocks in cluster mergers (see Giovannini \& Feretti
2002, and references therein).

\begin{figure}
\includegraphics[height=14pc]{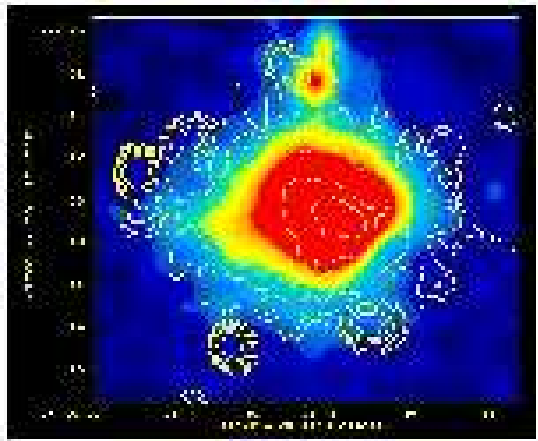}
\caption
{The cluster A2163 at z = 0.203 in radio and X-rays.  The contours
represent the radio emission in A2163 at 20 cm, showing an extended
radio halo (from Feretti et al. 2001).  The color scale represents the
ROSAT X-ray emission. The extended irregular X-ray structure indicates
the presence of a recent cluster merger.  The radio halo is one of the
most powerful and extended halos known so far. It shows a regular
shape, slightly elongated in the E-W direction.
\label{a2163rx}
}
\end{figure}

\begin{figure}
\includegraphics[height=18pc]{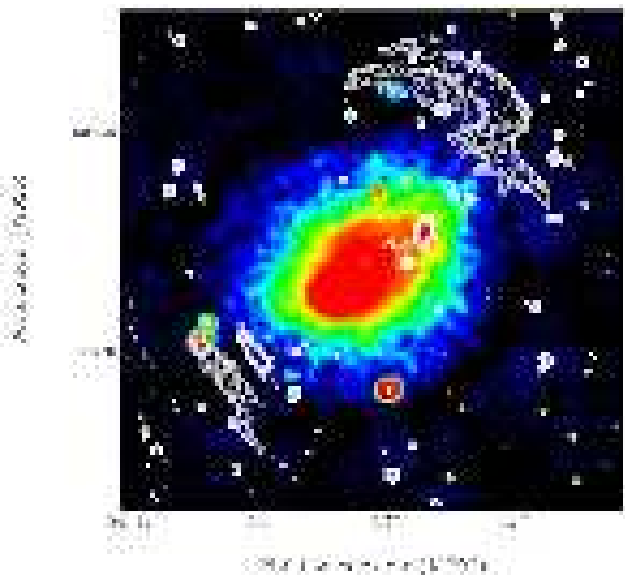}
\caption
{The cluster A3667 at z = 0.055 in radio and X-rays.  The contours
represent the radio emission at 843 MHz (from R\"ottgering et
al. 1997). The color scale represents the ROSAT X-ray 
emission. Two radio relics are located on opposite sides of the
cluster along the axis of the merger, with the individual radio
structures elongated perpendicularly to this axis.
\label{a3667rx}
}
\end{figure}

The Coma cluster is the first cluster where a radio halo (Coma C) and
a relic (1253+275) have been detected (Willson 1970, Ballarati et
al. 1981).  A typical example of a cluster radio halo is shown in
Fig. \ref{a2163rx}.  The radio halos are permeating the cluster
central regions, with typical extent of \gtsim~1 Mpc and steep
spectrum.  Limits of a few percent to the polarized emission have been
derived.  Relic sources are similar to halos in their low surface
brightness, large size and steep spectrum, but they are typically
found in the cluster peripheral regions (see Fig.  \ref{a3667rx}).
Unlike halos, relics are highly polarized ($\sim$ 20\%).

The formation and evolution of halo sources is still under
debate. Several suggestions were made for the mechanism transferring
energy into the relativistic electron population and for the origin of
relativistic electrons themselves.  Current models have been reviewed
by \citet{bruntaiw}. The relativistic particles could be injected in
the cluster volume by AGN activity (quasars, radio galaxies, etc.), or
by star formation in normal galaxies (supernovae, galactic winds,
etc).  Most of the particle production has occurred in the past and is
therefore connected to the dynamical history of the clusters.  This
population of {\it primary electrons} needs to be reaccelerated
(Brunetti et al. 2001, Petrosian 2001) to compensate for the radiative
losses. A recent cluster merger is the most likely process acting in
the reacceleration of relativistic particles.

Another class of models for the radiating particles in halos involves
{\it secondary electrons}, resulting from inelastic nuclear collisions
between the relativistic protons and the thermal ions of the ambient
intracluster medium.  The protons diffuse on large scale because their
energy losses are negligible.  They can continuously produce 
 in situ electrons, distributed through the cluster volume 
(Blasi \& Colafrancesco 1999, Miniati et al. 2001).

Different models have been suggested for the origin of the
relativistic electrons radiating in the relics.  There is increasing
evidence that the relics are tracers of shock waves in merger
events. This is consistent with their elongated structure almost
perpendicular to the merger axis (Fig. \ref{a3667rx}).  Active radio
galaxies may fill large volumes in the ICM with radio plasma, which
becomes rapidly invisible to radio telescopes because of radiation
losses of the relativistic electrons.  These patches of fossil radio
plasma are revived by adiabatic compression in a shock wave produced
in the ICM by the flows of cosmological large-scale structure
formation (En{\ss}lin et al. 1998, En{\ss}lin \& Gopal-Krishna 2001).

\begin{figure}
\includegraphics[scale=0.3]{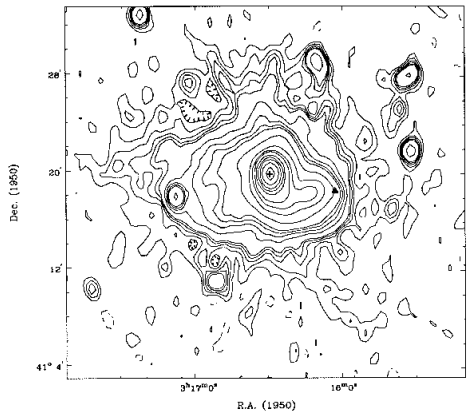}
\caption
{Radio contour map of the mini halo in the Perseus 
cluster, obtained at 92 cm with the WSRT at a resolution of
51\arcsec $\times$ 77\arcsec (RA$\times$DEC). 
The cross indicates the position of NGC1275, 
the triangle marks the position of NGC1272. The mini--halo
size in this image is $\sim$ 25\arcmin. This image is
from \citet{sijbr}.
\label{minih}
}
\end{figure}

For completeness, we wish to mention also another class of diffuse
radio sources associated with the ICM, the {\it mini-halos}. They are
detected around a dominant powerful radio galaxy at the center of
cooling core clusters, and have a total size of the order of $\sim$ 500
kpc, as detected in the Perseus cluster (Fig. \ref{minih}).  Unlike
radio halos and relics, mini-halos are typically found at the centers
of cooling core clusters and thus are not connected to recent cluster
mergers.  Although these sources are generally surrounding a powerful
central radio galaxy, it has been argued (Gitti, Brunetti \& Setti
2002, Gitti this volume) 
that the energetics necessary to their maintenance is not
supplied by the radio galaxy itself, but the electrons are
reaccelerated by MHD turbulence in the cooling core region. The
possibility of a hadronic origin of the relativistic electrons from
the interaction of cosmic ray protons with the ambient thermal protons
has been suggested by \citet{pfromm}.

The observational properties of radio halos and relics
are presented here, with emphasis on the information that is
derived by the comparison between the radio and X-ray emission.
The intrinsic parameters quoted in this paper are computed with a
Hubble constant H$_0$ = 50 km s$^{-1}$ Mpc$^{-1}$ and a deceleration
parameter q$_0$ = 0.5.

   \begin{figure}
   \centering
   \includegraphics[width=7cm]{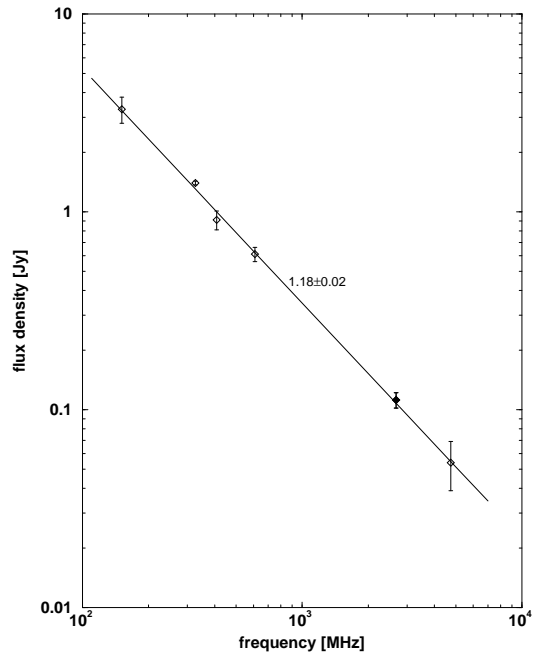}
      \caption{Total radio spectum of the Coma relic 1253+275
(from Thierbach, Klein \& Wielebinski 2003).
              }
         \label{comarelicsp}
   \end{figure}

   \begin{figure}
   \centering
   \includegraphics[width=7cm]{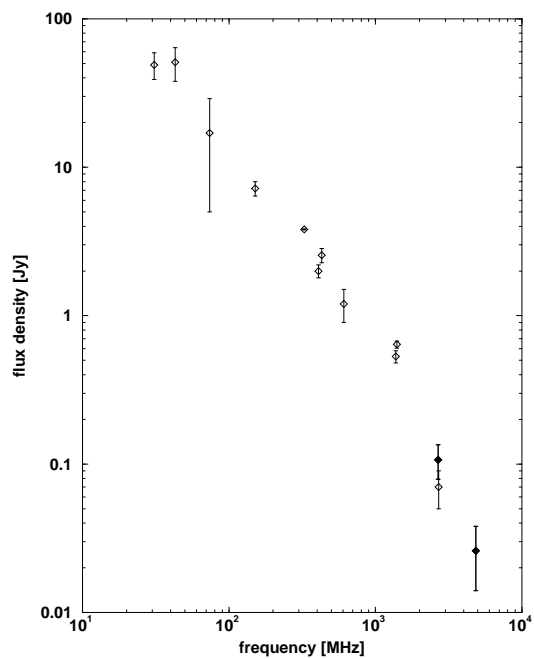}
      \caption{Total radio spectrum of the radio halo Coma C
(from Thierbach, Klein \& Wielebinski 2003).
              }
         \label{comacsp}
   \end{figure}

\section{What do we know from the radio data}

The typical radio powers of halo and relic sources are of the order of
10$^{24}$ -- 10$^{25}$ W Hz$^{-1}$ at 1.4 GHz.  The minimum energy
densities in diffuse sources, computed with standard assumptions, are
between $\sim$ 5 10$^{-14}$ and 5 10$^{-13}$ erg cm$^{-3}$.
The equipartition magnetic fields are of the order of $\sim$ 0.1 -- 1
$\mu$G.

Important information on the physical conditions in the radio sources
is obtained from the radio spectra, which reflect the energy
distribution of the radiating electrons.

   \begin{figure}
   \centering
\includegraphics[bb=18 144 400 500,scale=0.6,clip]{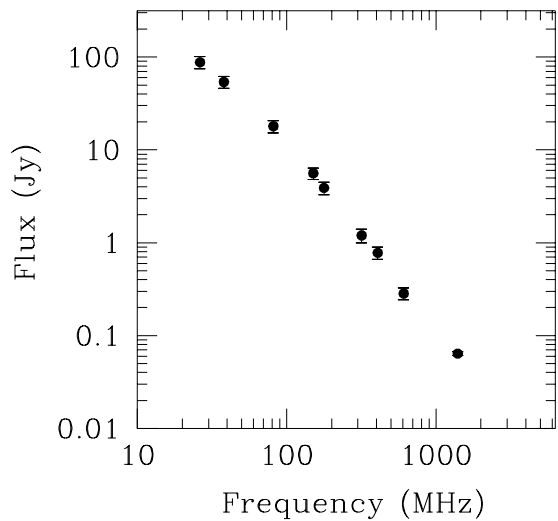}
      \caption{Integrated spectrum of the diffuse source in A1914
(Bacchi et al. 2003).
              }
         \label{a1914sp}
   \end{figure}

The spectra of halos and relics are steep, typical of aged radio
sources ($\alpha$ \gtsim~ 1\footnote {$S_{\nu}\propto \nu ^{-\alpha}$
through this paper}). As an example, the spectrum of the relic in the
Coma cluster has a spectral index = 1.18 (Fig. \ref{comarelicsp}).
The spectral index between 0.3 and 1.4 GHz of the radio halo in A665 
is $\alpha^{1.4}_{0.3}$ = 1.04. In the cluster A2163, shown in
Fig. \ref{a2163rx}, the spectral index between the same frequencies is
$\alpha^{1.4}_{0.3}$ = 1.18 (Feretti et al.  2004).

The spectrum of the Coma cluster halo is characterized by a steepening
at high frequencies, which has been recently confirmed by single dish
data (Fig. \ref{comacsp}). The spectrum of the radio halo in A1914
(Fig. \ref{a1914sp}) is very steep, with an overall slope of $\alpha
\sim$ 1.8. A possible high frequency curvature is discussed by
\citet{komiss}.  In A754, \citet{bacchi} estimate
$\alpha_{0.07}^{0.3}$ $\sim$ 1.1, and $\alpha_{0.3}^{1.4}$ $\sim$ 1.5,
and infer the presence of a possible spectral cutoff.  Indication of a
high frequency spectral steepening is also obtained in the halo of
A2319, where \citet{fer2319} report $\alpha_{0.4}^{0.6}$ $\sim$ 0.9
and $\alpha_{0.6}^{1.4}$ $\sim$ 2.2.

In general, from the spectra of diffuse radio sources, it is derived
that the radiative lifetime of the relativistic electrons, considering
synchrotron and inverse Compton energy losses, is of the order of
$\sim$ 10$^8$ yr.  This is too short to allow the particle diffusion
throughout the cluster volume. Thus, the radiating electrons cannot
have been produced at some localized point of the cluster, but they
must undergo {\it in situ} energization.

\section{What do we know from the radio - X-ray comparison: relativistic plasma and cluster merger}

The arguments in favor of a connection between the presence of radio
halos and the existence of merger processes have been reported in
several papers (see e.g. Feretti 2003, and references therein).  
Studies at the X-ray energies are of major
importance in this respect since the thermal gas, which is directly
observed in X-rays, bears the most evident signatures of cluster
mergers.  Complementary information on the cluster evolution can also
be obtained by optical data (Girardi \& Biviano 2002), as shown by 
the recent analysis of the halo cluster A2219 presented by
\citet{boschin}.

In all clusters with halos and relics, substructures, distortions in
the brightness distribution, temperature gradients and gas shocks are
detected, which can be interpreted as the result of strong dynamical
activity and sub-clump interaction.  Cluster mergers are among the
most energetic phenomena in the Universe, releasing gravitational
binding energies of about 10$^{64}$ erg. They generate shocks, bulk
flows, turbulence in the ICM. These processes would provide energy to
reaccelerate the radiating particles all around the cluster.

The connection to cluster mergers would explain at least in part why
the diffuse emission is not detected in all clusters. 
Unlike the thermal X-ray
emission, the presence of diffuse radio emission is not common in
clusters of galaxies.  In a complete sample, 5\% of clusters have a
radio halo source and 6\% have a peripheral relic source 
(at the NRAO VLA Sky Survey surface brightness limit, e.g. Giovannini
\& Feretti 2002).  The detection rate of diffuse radio sources
increases with the cluster X-ray luminosity, reaching $\sim$ 35\% in
clusters with X-ray luminosity larger than $\sim$ 10$^{45}$ erg
s$^{-1}$.  We note, however, that not all clusters showing recent
mergers do exhibit a halo and/or a relic. A peculiar example is given
by the cluster A3667, presented in Fig. \ref{a3667rx}, which shows two
opposite relics, but not a radio halo, despite of the existence of a
violent recent merger (Vikhlinin, Markevitch \& Murray 2001). This
case, and similar cases, will be discussed in Sect. 3.3.

\subsection{Radio structure vs X-ray structure and gas temperature}

A close similarity between the radio halo structure and the X-ray cluster
structure detected by the ROSAT satellite was first noted by
\citet{deiss} for the Coma cluster, and quantitatively confirmed in
Coma and in other clusters by \citet{govenssl}.  This similarity
indicates a close link between the physical conditions of the radio
source and those of the thermal component.  Since the structure of the
X-ray emission is generally related to a cluster merger process, a
close connection between the structure of the halo and that of the
X-ray gas supports a connection between the halo radio emission and
the merger.  A Chandra observation of A2744 (Kempner \& David 2004)
shows that the strong correlation between the radio and X-ray surface
brightness found in the lower resolution ROSAT data is also visible
at high resolution. In addition, Chandra data reveals that the main
cluster is in a highly disturbed state, with several shocks in the
cluster core. Also, there is a small merging subcluster, which
is being ram pressure stripped by its interaction with the main cluster.

High resolution Chandra X-ray data have been recently analyzed for a
sample of clusters with halos or relics.  In some clusters there is a
correlation between the radio halo emission and the hot gas regions
(Govoni et al. 2004). In all these clusters temperature gradients and
gas shocks are detected confirming the presence of mergers (Govoni et
al. 2004; Markevitch, this volume).  Although it may be difficult in
several case to disentangle the geometry of the cluster merger, it is
generally deduced that the cluster merger is likely to supply the
energy for the electron reacceleration.  Velocities inferred for
merger shocks at the cluster centers have Mach numbers of the order of
$\sim$ 1-2 (Markevitch, Vikhlinin \& Forman 2003, and references
therein), which seem too low to accelerate the radio halo electrons
(Gabici \& Blasi 2003).  This lends support to turbulent
reacceleration models for radio halo formation.

High resolution X-ray data of the Coma cluster have been obtained with
XMM-Newton. There evidence of recent merger activity at scales larger
than 10\arcmin, whereas the cluster core is suggested to be in a
basically relaxed state (Arnaud et al. 2001). 

\subsection{Radio spectral index maps vs X-ray emission}

Spectral index maps represent a powerful tool to study the properties
of the relativistic electrons and of the magnetic field, and to
investigate the connection between the electron energy distribution
and the ICM.  By combining high resolution spectral information and
X--ray images it is possible to study the thermal--relativistic plasma
connection both on small scales (e.g. spectral index variations
versus clumps in the ICM distribution) and larger scales (e.g. radial
spectral index trends).

The first spectral index image of a radio halo has been obtained by
\citet{giovanninicoma} for Coma C, using data at 327 MHz from the
Westerbork Synthesis Radio Telescope (WSRT) and data at 1.4 GHz from
the VLA and the Dominion Radio Astronomy Observatory. The image shows
a flat spectrum in the center ($\alpha\simeq$ 0.8) and a progressive
steepening with increasing distance from the center (up to
$\alpha\simeq$ 1.8 at a distance of about 15\arcmin).  This trend
is confirmed by a new spectral index map derived by comparing the 1.4
GHz image obtained with the Effelsberg single dish by \citet{deiss},
and the 327 MHz image obtained from the combination of VLA and WSRT
data (Giovannini et al. 2003).  The high sensitivity of the images
allows the computation of the spectral index up to $\sim$ 30\arcmin~
from the cluster center, where the spectral index is $\alpha\simeq$ 2.

Since the diffusion velocity of relativistic particles is low with
respect to their radiative lifetime, the radial spectral steepening
cannot be simply due to ageing of radio emitting electrons. Therefore
the spectral steepening must be related to the intrinsic evolution of
the local electron spectrum and to the radial profile of the cluster
magnetic field.
It has been shown by \citet{brunetti} that a relatively general
expectation of models invoking reacceleration of relic particles is a
radial spectral steepening in the synchrotron emission from radio
halos.  The steepening, that is difficult to reproduce by other models
such as those invoking secondary electron populations, is due to the
combined effect of a radial decrease of the cluster magnetic field
strength and the presence of a high energy break in the energy
distribution of the reaccelerated electron population.  In the
framework of reacceleration models the radio spectral index map can be
used to derive the physical conditions prevailing in the clusters,
i.e. reacceleration coefficients (efficiency) and magnetic field
strengths. From the application of this method to Coma C,
\citet{brunetti} obtained large 
scale reacceleration efficiencies of the
order of $\sim 10^{8}$ yr$^{-1}$ and magnetic field strengths ranging
from 1--3 $\mu$G in the central regions down to 0.05--0.1 $\mu$G in the
cluster periphery.

Maps of the radio spectral index between 0.3 GHz and 1.4 GHz have been
very recently obtained for two more radio halos, those in A665 and
A2163, using VLA data with an angular resolution of the order of
$\sim$ 1\arcmin~ (Feretti et al. 2004).

\begin{figure}
\includegraphics[width=7cm]{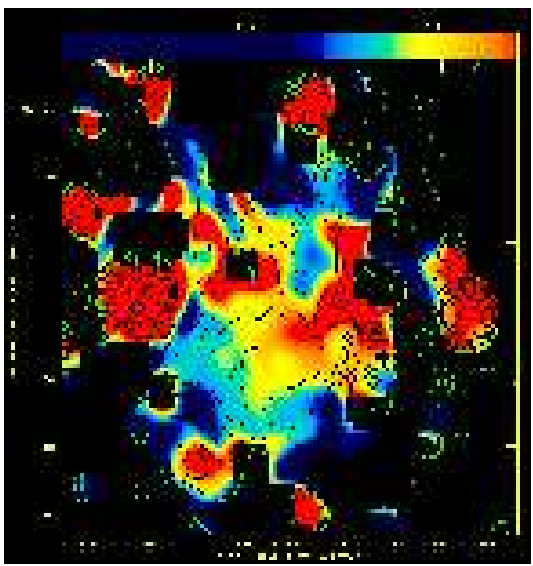}
\caption{ Color-scale image of the spectral index of A665 (at z =
0.1818) between 0.3 GHz and 1.4 GHz. The contours indicate the radio
emission at 1.4 GHz (from Giovannini \& Feretti 2000).  The radio
emission is asymmetric with respect to the cluster center. It is
brighter and more extended toward NW, which is the region of flatter
spectrum.
\label{a665spix}
}
\end{figure}

\begin{figure}
\includegraphics[width=8.5 cm]{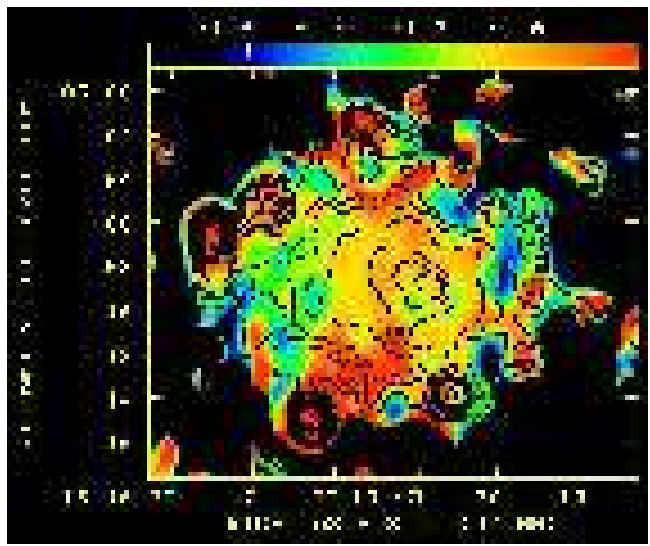}
\caption{ Color-scale image of the spectral index of A2163 (at z =
0.203) between 0.3 GHz and 1.4 GHz.  The contours indicate the radio
emission at 1.4 GHz (from Feretti et al. 2001).  The spectrum is
flatter in the vertical region across the cluster center.
\label{a2163spix}
}
\end{figure}

The spectral index map in A665 is clumpy (Fig. \ref{a665spix}).  The
spectrum in the central halo region is rather constant, with spectral
index values between 0.8 and 1.2 within one core radius from the
cluster center (i.e. within $\sim$ 95\arcsec).  In the northern region
of lower radio brightness, the spectrum is flatter.  This is the
region where asymmetric extended X-ray emission is present, indicating
the existence of a recent major merger.  The gas
temperature here (Markevitch \& Vikhlinin 2001, Govoni et al. 2004)
shows strong variations, from about 12 keV in the N-E to about 8 keV
in the S-W. Therefore it seems that this region, which is presently
strongly influenced by the merger, is a shocked area where the gas at
different temperatures is still in the process of mixing.  In the
southern cluster region the spectrum steepens significantly from the
center to the periphery, with the spectral index gradually increasing
from $\alpha$ $\sim$ 1 to $\alpha$ \gtsim~2 at a distance of about
4\arcmin. In the southern bright edge of the radio halo the presence
of a hot shock has been revealed by  Chandra X-ray data (Markevitch
\& Vikhlinin 2001). No significant spectral flattening is detected at this
position.  This is consistent with the fact that shocks in major
mergers are too weak for particle acceleration (Gabici \& Blasi 2003)
and indeed the Mach number of the shock in A665 is $\sim$ 2
(Markevitch \& Vikhlinin 2001). Our result supports the scenario that
cluster turbulence might be the major responsible for the supply of
energy to the radiating electrons.

The spectral index map of A2163 is rather constant in the central
region with the spectral index values between 1 and 1.1.  On larger scales
it is clumpy with the evidence that the western halo region has a
flatter spectrum than the eastern region. Particularly, there is a vertical region
crossing the cluster center and showing flatter spectrum, with a clear
evidence of spectral flattening both at the northern and 
southern halo boundaries.  The X-ray data show a complex
morphology and temperature distribution indicating that the cluster
central region is in a state of violent motion (Markevitch \&
Vikhlinin 2001, Govoni et al. 2004).  The N-S extent of the region
with flat spectrum is in support of a merger occurring in the E-W
direction, as indicated by the X-ray brightness distribution. The
complexity of the merger is reflected in the complexity of
the spectral index map.

The spectral index maps presented above show features (flattening and
patches) which are indicative of a complex shape of the radiating
electron spectrum, and are therefore in support of electron
reacceleration models.  Regions of flatter spectrum are found to be
related to the recent merger activity in these clusters.  This is the
first strong confirmation that the cluster merger supplies energy to
the radio halo.  We estimated that the energy injected into the
electron population in the region directly influenced by the merger is
larger by a factor of $\sim$ 2.5.  Alternatively, if electrons have
been reaccelerated in the past and they are simply ageing, the
flatter spectrum would reflect a spectral cutoff at higher energies. We
estimate that the electrons in the flat spectrum regions have been
reaccelerated more recently by $\sim$ 5 10$^7$ yr.

In the undisturbed cluster regions the spectrum steepens with the
distance from the cluster center.  This is interpreted as the result
of the combination of the magnetic field profile with the spatial
distribution of the reacceleration efficiency, thus allowing us to set
constraints on the radial profile of the cluster magnetic field
(see Feretti et al. 2004 for the details).

   \begin{figure}
   \centering
   \includegraphics[bb=18 144 480 615,scale=0.5,clip]{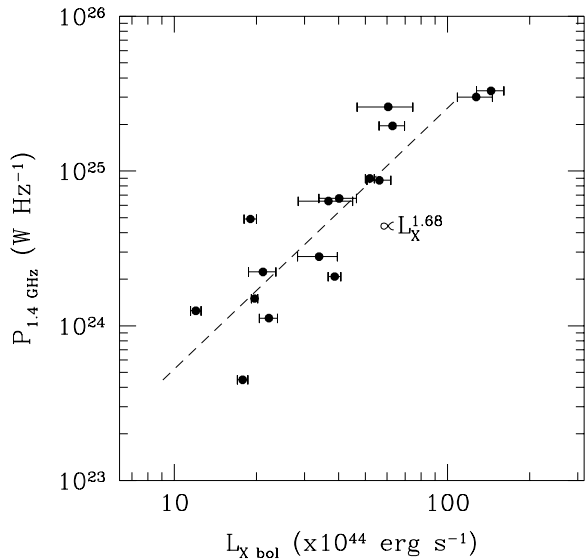}
      \caption{ Monochromatic radio power at 1.4 GHz of halos larger
than 1 Mpc versus cluster bolometric X-ray luminosity. The dashed line
represents the best fit to the data.  }
        \label{powerlum}
   \end{figure}

A study of the spectral index distribution in the relics of A3667 has
been performed by \citet{melanie}. A steepening from the external to
the internal rim, consistent with electron reacceleration in the
merger shock, is found.

\subsection{Radio quantities vs X-ray quantities}

A correlation between the monochromatic radio power  of the halo 
at 1.4 GHz and the bolometric X-ray luminosity of the parent cluster
(Fig. \ref{powerlum}) was first noted by Liang et al. (2000) and
confirmed by later studies (see e.g. Giovannini \& Feretti 2002).
Using only clusters with giant radio halos (size $>$ 1 Mpc) the best
fit between radio and X-ray luminosity is P$_{\rm 1.4 GHz} \propto
L_{\rm X}^{1.68\pm0.15}$.  An overall correlation is still present,
although with a larger dispersion, if halos of smaller extent and/or
relics are added.

We stress that this correlation is only valid for clusters showing
radio halos, i.e. merging clusters, thus it cannot be generalized to
all clusters of galaxies.  It is not clear if it can be extrapolated
to low radio powers and low X-ray luminosities. Moreover, it is difficult to
extend such a correlation by including the upper limits for the
undetected halos, since the computation of an upper limit to the radio
power would imply the knowledge of the total size of a possible radio
halo.

\begin{figure}
   \centering
   \includegraphics[bb=18 144 480 615,scale=0.5,clip]{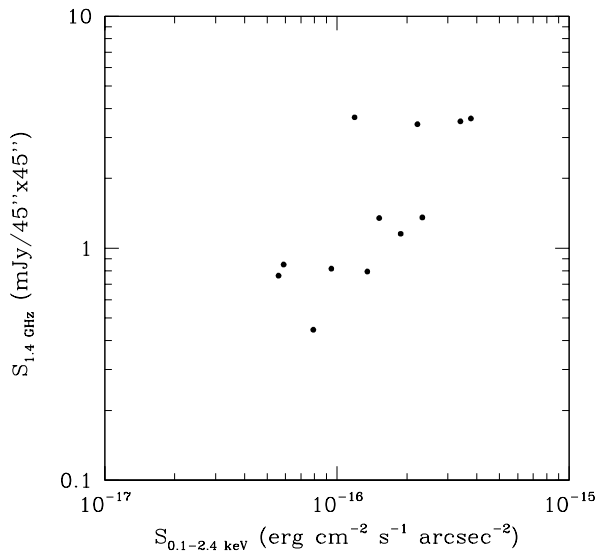}
\caption{ Radio brightness at 1.4 GHz in mJy/(45\arcsec$\times$45\arcsec) versus
cluster X-ray brightness in the ROSAT band.  }
         \label{brbx}
   \end{figure}

Fig. \ref{brbx} shows the radio surface brightness of the halo at 1.4
GHz versus the X-ray surface brightness, which are two directly
observable parameters. The radio brightness is the average over a
cluster central region within one cluster core radius. For
convenience, it is expressed in the same units as the radio brightness
of the NRAO VLA Sky Survey images, which are obtained with an
observing beam of 45\arcsec. The X-ray brightness is the average over
the same region as the radio brightness, and has been obtained using
ROSAT X-ray images, thus it refers to the ROSAT energy band 0.1 - 2.4
keV. This plot contains a limited number of clusters, because of the
difficulty in deriving the X-ray surface brightness.

\begin{figure}
   \centering
   \includegraphics[bb=18 144 480 615,scale=0.5,clip]{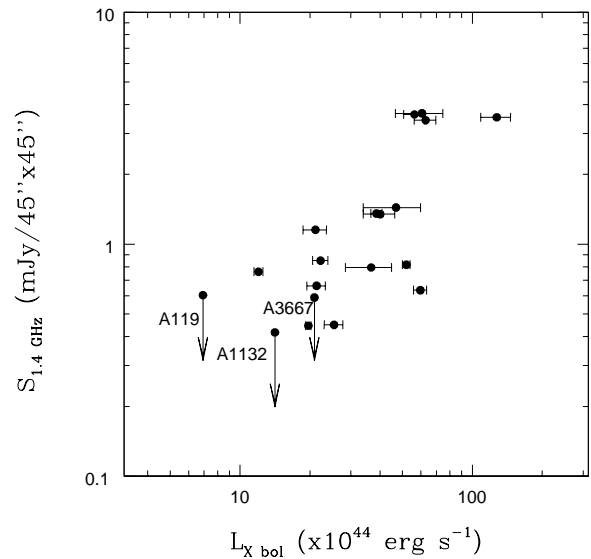}
      \caption{ Radio brightness at 1.4 GHz in mJy/(45\arcsec$\times$45\arcsec)
versus cluster bolometric X-ray luminosity. Upper limits to the radio brightness
of a possible halo are given for the merging clusters A119, A1132 and A3667. }
         \label{brlxl}
   \end{figure}

Fig. \ref{brlxl} presents the radio surface brightness computed as
above versus the clusters bolometric X-ray luminosity. This choice of
parameters is convenient because the cluster luminosity is available
for a large number of clusters, observed by different X-ray
satellites.  This correlation can be used to set upper limits to the
radio emission for those clusters where a radio halo is not detected.
One of these clusters is A3667, which hosts two radio relics, but no
central radio halo.  We report also limits for two other merging
clusters observed by \citet{giova665}.  It is interesting to note that
these upper limits are consistent with the correlation, thus
indicating that merging clusters with low X-Ray luminosity might host
faint radio halos which could only be detected by very sensitive radio
observations possibly with future generation instruments (LOFAR,
SKA). Future data will indeed clarify if all merging clusters
host a radio halo.

\section{Radio - X-ray comparison: non-thermal emission}

X-ray emission of non-thermal origin is expected in clusters with
diffuse radio sources, as the high energy relativistic electrons
($\gamma$ $\sim$ 10$^4$) scatter off the cosmic microwave background,
boosting photons from this radiation field to the hard X-ray domain by
inverse Compton (IC) process.  Since the X-ray and radio emissions are
produced by the same population of electrons undergoing inverse
Compton and synchrotron energy losses, respectively, the ratio between
the X-ray and the radio luminosities is proportional to the ratio
between the CMB and the magnetic field energy densities.  Thus the
comparison between radio and hard X-ray emission enables the
determination of the electron density and of the mean magnetic field
directly, without invoking equipartition.

A significant breakthrough in the measurement of hard X-ray emission
was recently obtained owing to the improved sensitivity and wide
spectral capabilities of the BeppoSAX and the Rossi X-ray Timing
Explorer (RXTE) satellites (see the review by Fusco-Femiano et
al. 2003, and references therein). The detection of significant
non-thermal hard X-ray emission at energies \gtsim~ 20 keV has been
reported for the two clusters Coma and A2256.  From a recent 300 ksec
exposure on Coma, obtained with BeppoSAX, the detection in this
cluster was confirmed and strengthened by \citet{fuscocoma}. On
the other hand, \citet{molendi} did not confirm the non thermal X-ray
detection.  However, they analyzed the data using a different software
probably less sensitive to faint emission.

The 20-80 keV flux in Coma is $\sim$1.5 10$^{-11}$ erg cm$^{-2}$
s$^{-1}$, which leads to a magnetic field of $\sim$ 0.2 $\mu$G.  In
A2256, the flux in the same energy range is $\sim$9 10$^{-12}$ erg
cm$^{-2}$ s$^{-1}$. A magnetic field of $\sim$ 0.05 $\mu$G is derived
for the northern cluster region, where the radio relic is
detected, while a higher field value, $\sim$0.5 $\mu$G, could be
present at the cluster center, in the region of the radio halo.

A detection has been obtained for A754, but the presence of point
sources in the field of view makes it unlikely the IC interpretation.
For the clusters A119, A2163, A2199 and A3667, only upper limits to
the non-thermal X-ray emission have been derived.  A possible
detection of localized IC emission associated with the radio relic and
with merger shocks has been claimed in A85 from ROSAT data (Bagchi,
Pislar \& Lima-Neto 1998).

BeppoSAX ceased its activity in April 2002.  Future studies of non
thermal X-ray emission in clusters will be possible with INTEGRAL,
whereas the satellite XMM-Newton will be suitable for the
investigation of clusters of low temperature.

\section{Conclusions}

Massive clusters of galaxies showing strong dynamical activity and
merger processes can host diffuse radio emission, which demonstrates
the existence of relativistic particles and magnetic fields in the
ICM.

From the comparison between radio and X-ray emission there is evidence
that recent merger phenomena would provide the energy for the
relativistic electron reacceleration, thus allowing the production of
a detectable diffuse radio emission.

The connection between halos and mergers favors halo models where the
radiating particles are primary electrons reaccelerated in situ.  The
spectral index properties in radio halos (high frequency cutoff and radial
steepening) can be easily reproduced by models invoking the
reacceleration of the relativistic particles.

Spectral index maps of the halos of A665 and A2163 show regions of
flatter spectrum that appear to trace the geometry of recent merger
activity as suggested by X-ray maps.  No evidence of spectral
flattening at the location of the hot shock detected in A665 is found,
favoring the scenario that cluster turbulence might be the major
responsible for the electron reacceleration. This scenario is also
supported by the relatively low Mach numbers of shocks at the cluster
centers.

It seems that ongoing violent mergers may play a crucial role in
determining the conditions of the radiating particles and of the
magnetic fields in clusters. A question which is still unanswered is
whether all merging clusters have cluster-wide radio halos. This
will be answered by systematic deep studies with future instruments.

\section*{Acknowledgments}

I wish to thank the organizers for making possible this enjoyable and
scientifically profitable conference.  I am grateful to my
collaborators G. Brunetti, R. Fusco-Femiano, I.M. Gioia,
G. Giovannini, and G. Setti for many interesting discussions.  This
work was partially funded by the Italian Space Agency.

\end{document}